\documentclass[12pt]{article}
\usepackage{html}
\usepackage{epsf}
\usepackage[pdftex]{graphicx}
\usepackage{graphicx}
\usepackage{amssymb}
\usepackage{amsmath}
\usepackage{hyperref}
\begin{document}
\begin{titlepage}
\includegraphics[width=150mm]{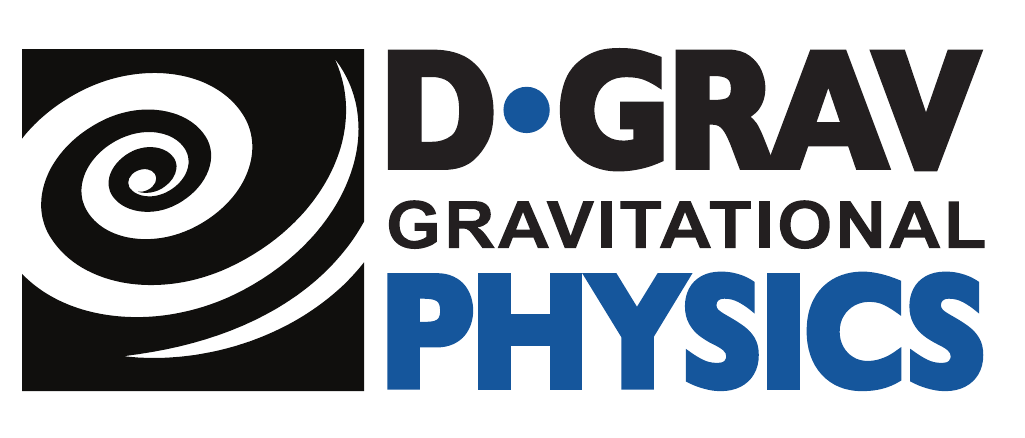}
\begin{center}
{ \Large {\bf MATTERS OF GRAVITY}}\\ 
\bigskip
\hrule
\medskip
{The newsletter of the Division of Gravitational Physics of the American Physical 
Society}\\
\medskip
{\bf Number 51 \hfill June 2018}
\end{center}
\begin{flushleft}
\tableofcontents
\end{flushleft}
\end{titlepage}
\vfill\eject
\begin{flushleft}
\section*{\noindent  Editor\hfill}
David Garfinkle\\
\smallskip
Department of Physics
Oakland University
Rochester, MI 48309\\
Phone: (248) 370-3411\\
Internet: 
\htmladdnormallink{\protect {\tt{garfinkl-at-oakland.edu}}}
{mailto:garfinkl@oakland.edu}\\
WWW: \htmladdnormallink
{\protect {\tt{http://www.oakland.edu/physics/Faculty/david-garfinkle}}}
{http://www.oakland.edu/physics/Faculty/david-garfinkle}\\

\section*{\noindent  Associate Editor\hfill}
Greg Comer\\
\smallskip
Department of Physics and Center for Fluids at All Scales,\\
St. Louis University,
St. Louis, MO 63103\\
Phone: (314) 977-8432\\
Internet:
\htmladdnormallink{\protect {\tt{comergl-at-slu.edu}}}
{mailto:comergl@slu.edu}\\
WWW: \htmladdnormallink{\protect {\tt{http://www.slu.edu/arts-and-sciences/physics/faculty/comer-greg.php}}}
{http://www.slu.edu/arts-and-sciences/physics/faculty/comer-greg.php}\\
\bigskip
\hfill ISSN: 1527-3431


\bigskip

DISCLAIMER: The opinions expressed in the articles of this newsletter represent
the views of the authors and are not necessarily the views of APS.
The articles in this newsletter are not peer reviewed.

\begin{rawhtml}
<P>
<BR><HR><P>
\end{rawhtml}
\end{flushleft}
\pagebreak
\section*{Editorial}

The next newsletter is due December 2018.  Issues {\bf 28-51} are available on the web at
\htmladdnormallink 
{\protect {\tt {https://files.oakland.edu/users/garfinkl/web/mog/}}}
{https://files.oakland.edu/users/garfinkl/web/mog/} 
All issues before number {\bf 28} are available at
\htmladdnormallink {\protect {\tt {http://www.phys.lsu.edu/mog}}}
{http://www.phys.lsu.edu/mog}

Any ideas for topics
that should be covered by the newsletter should be emailed to me, or 
Greg Comer, or
the relevant correspondent.  Any comments/questions/complaints
about the newsletter should be emailed to me.

A hardcopy of the newsletter is distributed free of charge to the
members of the APS Division of Gravitational Physics upon request (the
default distribution form is via the web) to the secretary of the
Division.  It is considered a lack of etiquette to ask me to mail
you hard copies of the newsletter unless you have exhausted all your
resources to get your copy otherwise.

\hfill David Garfinkle 

\bigbreak

\vspace{-0.8cm}
\parskip=0pt
\section*{Correspondents of Matters of Gravity}
\begin{itemize}
\setlength{\itemsep}{-5pt}
\setlength{\parsep}{0pt}
\item Daniel Holz: Relativistic Astrophysics,
\item Bei-Lok Hu: Quantum Cosmology and Related Topics
\item Veronika Hubeny: String Theory
\item Pedro Marronetti: News from NSF
\item Luis Lehner: Numerical Relativity
\item Jim Isenberg: Mathematical Relativity
\item Katherine Freese: Cosmology
\item Lee Smolin: Quantum Gravity
\item Cliff Will: Confrontation of Theory with Experiment
\item Peter Bender: Space Experiments
\item Jens Gundlach: Laboratory Experiments
\item Warren Johnson: Resonant Mass Gravitational Wave Detectors
\item David Shoemaker: LIGO 
\item Stan Whitcomb: Gravitational Wave detection
\item Peter Saulson and Jorge Pullin: former editors, correspondents at large.
\end{itemize}
\section*{Division of Gravitational Physics (DGRAV) Authorities}
Chair: Emanuele Berti; Chair-Elect: 
Gary Horowitz; Vice-Chair: Nicolas Yunes. 
Secretary-Treasurer: Geoffrey Lovelace; Past Chair:  Peter Shawhan; Councilor: Beverly Berger
Members-at-large:
Kelly Holley-Bockelmann, Leo Stein, Lisa Barsotti, Theodore Jacobson, Michael Lam, Jess McIver.
Student Members: Cody Messick, Belinda Cheeseboro.
\parskip=10pt

\vfill\eject

\section*{\centerline
{we hear that \dots}}
\addtocontents{toc}{\protect\medskip}
\addtocontents{toc}{\bf DGRAV News:}
\addcontentsline{toc}{subsubsection}{
\it we hear that \dots , by David Garfinkle}
\parskip=3pt
\begin{center}
David Garfinkle, Oakland University
\htmladdnormallink{garfinkl-at-oakland.edu}
{mailto:garfinkl@oakland.edu}
\end{center}

Vicky Kalogera has been elected to the National Academy of Sciences.

Nicolas Yunes has been elected Vice-Chair of DGRAV.  Jess McIver and Michael Lam have been elected Members-at-large of the DGRAV Executive Committee.  Belinda Cheesboro has been elected Student Student Member of the DGRAV Executive Committee.

Hearty Congratulations!

\vfill\eject
\section*{\centerline
{DGRAV Student Travel Grants}
}
\addtocontents{toc}{\protect\medskip}
\addcontentsline{toc}{subsubsection}{
\it DGRAV Student Travel Grants, by Beverly Berger}
\parskip=3pt
\begin{center}
Beverly Berger, LIGO
\htmladdnormallink{beverlyberger-at-me.com}
{mailto:beverlyberger@me.com}
\end{center}

In the year of the General Relativity Centennial, the APS Topical Group in Gravitation (now the Division of Gravitational Physics) began an effort to raise funds to support the travel of student members to present their work at the April Meeting. This travel grant program, started in 1999 with GGR (now DGRAV) operating funds, has been wildly successful. This year more than 40 students received such grants as part of the largest participation ever by DGRAV (or GGR) members in the April Meeting. 

The fund-raising program began because DGRAV operating funds cannot meet the demand from ever increasing numbers of applicants. Students are the future of our field. We would like to assist as many as possible to experience the excitement (and networking opportunities) of the April Meeting. 

Since this campaign began, DGRAV has raised \$50,000. Thank you, donors! However, the goal for the campaign is \$75,000. Please donate to help us reach this goal. If your students have benefited from the student travel grant program or you, when a student, received an award, or if you believe it is important to support students this way, please consider a donation.
 
As a special incentive, DGRAV is offering a General Relativity Centennial commemorative medallion, designed and minted for the 2015 Centennial banquet, to donors who request one to express our gratitude for your support. These are in limited quantity and will be distributed in order of the gift receipt date. 

For more information or to contribute on line, please visit the campaign website, 
\href{https://www.aps.org/about/support/campaigns/dgrav/index.cfm}{https://www.aps.org/about/support/campaigns/dgrav/index.cfm}, which also lists donors to the campaign.
To contribute by mail, please make your check payable to the American Physical Society, note ``DGRAV Travel'' in the memo, and mail to Irene I. Lukoff, Director of Development, One Physics Ellipse, College Park, MD 20740. 

If you would like a medallion, please send a request with your mailing address to Peter Shawhan, \htmladdnormallink{pshawhan@umd.edu}{mailto:pshawhan@umd.edu}, once your donation is received.

\vfill\eject
\section*{\centerline
{Town Hall Meeting}
}
\addtocontents{toc}{\protect\medskip}
\addcontentsline{toc}{subsubsection}{
\it Town Hall Meeting, by Emanuele Berti}
\parskip=3pt
\begin{center}
Emanuele Berti, University of Mississippi
\htmladdnormallink{eberti-at-olemiss.edu}
{mailto:eberti@olemiss.edu}
\end{center}

Town Hall Meeting on ``Gravitational Wave Theory and Simulations in the Era of Detections''

Gravitational wave astronomy is finally a reality. The first multi-messenger observation of a binary neutron star merger (GW170817) was one of the biggest science stories of 2017, and binary black hole detections are becoming routine. However there is still a lot of work needed to coordinate the efforts of instrumentalists, data analysts, astrophysicists and gravitational theorists.

Gravitational-wave theorists, numerical relativists, astrophysicists and LIGO/Virgo members should work together as a single community to make the best of the wealth of data that will be collected during LIGO/Virgo's O3 run. For this reason, Beverly Berger and Manuela Campanelli proposed to organize a Town Hall Meeting on ``Gravitational Wave Theory and Simulations in the Era of Detections.''

The Town Hall Meeting, sponsored by DGRAV and co-sponsored by DAP, took place immediately after the DGRAV and DAP Business Meetings on Monday, April 16 at the APS April Meeting in Columbus, OH. Slides from both meetings are available online at 

\href{https://dgrav.org/2018/04/17/slides-from-the-2018-business-and-town-hall-meetings/}{https://dgrav.org/2018/04/17/slides-from-the-2018-business-and-town-hall-meetings/}

Laura Cadonati (on behalf of the LIGO Scientific Collaboration) presented the anticipated evolution of noise power spectral densities for LIGO/Virgo, as well as current plans for KAGRA, in the next few years. Advanced LIGO (aLIGO) is expected to reach target sensitivity around 2019. The proposed A+ (if funded by NSF) could be operating by 2024. The expected range for a BNS (1.4+1.4 solar masses) would go from 96 Mpc (O2) to 173 Mpc (aLIGO) and 325 Mpc (A+). The expected range for a BBH (30+30 solar masses) would go from 983 Mpc (O2) to 1606 Mpc (aLIGO) and 2563 Mpc (A+).

Laura and Lisa Barsotti discussed current plans to update the Virgo detector. Finally, Laura presented a table with the current list of priorities from the LSC-Virgo White Paper on Gravitational Wave Data Analysis and Astrophysics:
\href{https://dcc.ligo.org/LIGO-T1700214/public}{https://dcc.ligo.org/LIGO-T1700214/public}

Laura's presentation was followed by 7 short presentations on astrophysics science targets and numerical relativity tools (or results) that could be relevant for O3.

Carl Rodriguez showed some results from simulations of BBH mergers in clusters, suggesting that such mergers could produce signals with nonzero eccentricity in the LIGO band.

Davide Gerosa pointed out the importance of measuring precessional effects (beyond the ``effective spin'' parameter) to extract astrophysics from the observations.

Ilias Cholis asked whether merging BHs in globular clusters could experience runaway collisions leading to formation of intermediate mass BHs; if so, LIGO observations could be used to derive limits on the occupation fraction of intermediate mass BHs in globular clusters.

Zach Etienne pointed out that Moore's Law is slowing (at least for CPU-based codes - a point raised in discussions by Maurice van Putten and Kai Staats, among others) and that more efficient numerical relativity algorithms are necessary. He also pointed out the importance of writing software documentation for the benefit of students and postdocs.

Roland Haas gave an introduction to the Einstein Toolkit and its applications.

Antonios Tsokaros presented results on sequences of spinning BNS which suggest that corotating sequences can have low spin ($< 0.3$) even for close binaries.

Milton Ruiz presented simulations of BNS in GRMHD which imply (for causal EOS) that the maximum mass of the remnant should be below ~2.16 solar masses.

After these presentations, Bangalore Sathyaprakash gave an introduction to current plans for ``Third Generation'' (3G) detectors on behalf of the GWIC 3G Committee and 3G Science Case Team. He made the argument that LIGO and Virgo both have facility-imposed limits on sensitivity (implying at most a factor ~3 improvement in strain sensitivity, with gravity gradient noise limiting sensitivity below 10 Hz). Therefore there is a compelling case to build detectors that can observe deeper into the cosmos, with facilities that will be good ~30-40 years after construction. One of the GWIC charges is to ``commission a study of ground-based gravitational wave science from the global scientific community, investigating potential science vs. architecture vs. network configuration vs. cost trade-offs.'' The GWIC subcommittee has constituted five 3G subcommittees: (1) Science Case Team (3G-SCT), (2) R\&D Coordination, (3) Governance, (4) Agency Interfacing, (5) Community Networking. The Science Case will be developed by an international consortium of scientists under the leadership of the 3G-SCT, which consists of 18 members. For more details, see

\href{https://gwic.ligo.org/3Gsubcomm/}{https://gwic.ligo.org/3Gsubcomm/}

The 3G Science Case consortium is open to anyone who wishes to contribute to the development of the science case for 3G. If you are interested, please send a one-page CV and research interests relevant to 3G to either B.S. Sathyaprakash (bss25@psu.edu) or Vicky Kalogera (vicky@northwestern.edu).

Emanuele Berti (on behalf of the Executive Committee of the APS Division of Gravitational Physics) coordinated a lively discussion on how theorists could help LVC members and vice versa.

The discussion concerned the following main points:

1) The astrophysics/GR communities need the full posteriors to optimize the science return of LVC observations. How quickly should the LVC release these posteriors? Most participants agreed that the answer to this question is complex.

Bangalore Sathyaprakash explained that careful data analysis is necessary before releasing posteriors. Jolien Creighton argued that it is important for the LVC to retain some core science in order to maintain a vibrant collaboration, which in turn is vital to the larger community. A discussion involving various people (Will Farr,  Emanuele Berti, Chad Galley, Davide Gerosa and many others) followed.

It was argued that the current model for open data release is tuned to the (urgent) needs of observational astronomers. However, astrophysicists and theorists could benefit from a quicker release of the bulk properties of detected events. For example, Chad Galley pointed out that in other fields it is common to release data sets of lower quality which can be improved over time.

As the number of detections increases, one may consider releasing estimates of masses/distance/effective spins and other ``simple'' properties of each event for use by astronomers and modelers, and then refining parameter estimation for more in-depth applications (such as tests of general relativity).

2) Emanuele Berti also proposed that the collaboration between theorists and LVC members could follow the particle-physics model in two ways:

(i) At CERN, experimental physicists work in close collaboration with a ``theory division.'' The level of involvement, access to the data, and authorship details for the members of this ``theory division'' could and should be discussed on a case-by-case basis. It would be healthy to start discussions of this possibility within the LVC. Similar discussions in the recent past led to a more active LVC involvement of numerical relativists.

(ii) A key reference in particle physics is the ``particle data group'', which is updated annually to reflect the state of the art in the knowledge of particle properties, physical constants, etcetera. Various people (including Leo Stein, Emanuele Berti, Vitor Cardoso, Pedro Ferreira and Thomas Sotiriou, among others) have been discussing the possibility of creating and maintaining a ``gravity data group'' that could list the best known limits on (say) the mass of the graviton, Lorentz violation in gravity etcetera, and current constraints on various proposed modifications of general relativity. This effort requires manpower to maintain a website (or a similar ``living'' resource) and a white paper - that could be circulated on the arXiv - explaining how to interpret the constraints and pointing to relevant references.

\vfill\eject
\section*{\centerline {News from the International Society}
\centerline {on General Relativity and Gravitation (ISGRG)}
}
\addtocontents{toc}{\protect\medskip}
\addcontentsline{toc}{subsubsection}{
\it GRG Society, by Eric Poisson}
\parskip=3pt
\begin{center}
Eric Poisson, President of ISGRG
\htmladdnormallink{epoisson-at-uoguelph.ca}
{mailto:epoisson@uoguelph.ca}
\end{center}

The International Society on General Relativity and Gravitation is the only explicitly international organization for general relativity, gravitational physics, and mathematical gravitation. It was formed under the auspices of the International Union of Pure and Applied Physics (IUPAP) as an Affiliated Commission. The Society's purposes are to promote the study of General Relativity and Gravitation and to exchange information in the interest of its members and the profession.  

Times have been exciting at the Society since the great success of GR21 at Columbia University.   

Probably the most enjoyable activity at the Society is the awarding of various prizes to meritorious scientists. First and foremost is the \href{http://www.isgrg.org/IUPAPprize.php}{Young Scientist Prize}, sponsored by IUPAP. In 2017 the Prize was given to Aron Wall, for his fundamental contributions to our understanding of gravitational entropy and the generalized second law of thermodynamics. In 2018 the Prize was awarded to Sam Gralla, for his exceptional and broadly varied contributions to general relativity and relativistic astrophysics. The Society also gives out the \href{http://www.isgrg.org/bergwheelprize.php}{Bergmann-Wheeler Thesis Prize} for an outstanding Ph.D.\ thesis in the broad area of quantum gravity, and the \href{http://www.isgrg.org/ehlersprize.php}{J\"urgen Ehlers Thesis Prize} for an outstanding Ph.D. thesis in mathematical and numerical relativity. Nominations for both prizes are now open for the 2019 instalments. And every three years the Society recognizes the leading figures of our field by awarding a number of \href{http://www.isgrg.org/fellows.php}{Fellowships}; nominations for the 2019 batch are also open.  

Plans are well underway for the next GR meeting, GR22, which will be held jointly with the Thirteenth Amaldi Conference on Gravitational Waves in Valencia, Spain, in July, 2019. Most of the plenary speakers and session chairs have been identified, and a lot of information is already available on the \href{http://www.gr22amaldi13.com}{conference website}. The Local Organizing Committee is chaired by Jos\'e Antonio Font, and the Scientific Organizing Committee is chaired by Vitor Cardoso. They have been working very hard, and this should be a most exciting conference, in a beautiful part of the world. 

The Society's executive structure has changed in the last few months. The Constitution was amended to increase the Executive Committee by one member, to turn it from a three-member team (President, Deputy-President, Secretary) to a four-member team (President, President-Elect, Past-President, Secretary). The motivation for this change is to have the President-Elect learn the ropes for three years before assuming the role of President. With the previous structure the President came in cold, with no prior experience within the Executive. As previously, the President becomes Past-President (formally Deputy-President) after a three-year term, but now with somewhat reduced responsibilities. I am happy to report that this change in the Constitution, which required a two-thirds majority for adoption, was approved by a vast majority of the membership. 

Another recent change, also approved by the members, is the possibility for members of a National Society such as DGRAV to join the Society with a substantial reduction in the membership fee. Thus, one-year and three-year membership dues are reduced by 20\% for members of a National Society, and the price of a life membership is reduced by whopping 30\%. 

The recent excitement with gravitational waves and the Nobel Prize (two of three laureates are ISGRG members!) is filling our gravitational hearts with much pride and joy. As President of the Society, I am in awe of these achievements, and welcome the enviable place that gravitation is now occupying in world science. For more information about the Society's activities, please visit our \href{http://www.isgrg.org}{website}.   
\vfill\eject

\section*{\centerline
{What's new in LIGO}}
\addtocontents{toc}{\protect\medskip}
\addtocontents{toc}{\bf Research Briefs:}
\addcontentsline{toc}{subsubsection}{
\it  What's new in LIGO, by David Shoemaker}
\parskip=3pt
\begin{center}
David Shoemaker, MIT 
\htmladdnormallink{dhs-at-mit.edu}
{mailto:dhs@mit.edu}
\end{center}

While LIGO did not write a contribution for the MOG for December 2017, a number of the articles there made reference to the first detection of a binary neutron start inspiral (GW10817) and some of the happy fallout. We want to complement that with more of a view from the inside of the events and then a sketch of what's coming in the next half-year as we prepare for the O3 observing run. 

By July 2017, as the end of O2 approached, planned for the end of August 2017, we could already assess the run as moderately successful; several clear binary black holes had been detected, and we had a solid body of data which could then be re-analyzed after some processing to regress out some excess noise sources, and with the promise of perhaps some additional events. We were very pleased that Advanced Virgo was making progress in its commissioning, and with a rapid increase in their sensitivity, they set 1 August as the goal to join LIGO in O2 observing. The two LIGO detectors had sensitivities comparable to the 2015-16 O1 run, although due probably to a point absorber on an optic, the Hanford (LHO) instrument had a reach for binary neutron stars of about 60 Mpc (SNR of 8, averaged over positions in the sky and polarizations); Livingston (LLO) had a sensitivity of some 100 Mpc. Virgo had just in late July achieved a reach of 26 Mpc after a breakthrough in commissioning, and had just developed enough of an infrastructure, knowledge of the calibration, and understanding of the noise properties to allow the data to be properly interpreted. We had a firm date of the end of August to complete O2 -- there was lots of reinstallation and commissioning planned for O3. Thus, roughly one month of common observation with Virgo appeared feasible. Based on the detection rates in O2 we had only modest hope that an event might be caught during that month. With that backdrop, any event would be considered a wonderful surprise. 

On August 14, the first surprise arrived: An unambiguous binary black hole signal triggered the pipelines watching the two LIGO instruments (the flagship detection software was not yet ready to read in the three instruments in real time), leading to the new detection GW170814 [PRL 119, 141101 (2017)]. The system, of 31 and 25 solar masses and at a distance of 540 Mpc, was seen after combining the signals from the three instruments with a false alarm rate of some 1/27,000 years. While the signal in Virgo was not large against the noise floor, the probability of the signal observed being due to chance was less than $3\times 10^{-3}$. From this detection, two major steps forward in our young observational field were possible. 

First, the localization of the source was significantly improved. The addition of Virgo enabled a much smaller angular (roughly a factor of 20 times smaller) area than could be determined from the two LIGO instruments alone, and the distance estimate also improved such that the volume determination for the source was reduced by a factor of 34. These data were sent out in the form of a first-ever 3-detector skymap as quickly as possible to the electromagnetic and particle observers with whom we had memoranda of understanding; no signal was identified that was likely associated with the coalescence. But the key feature of the network had been realized.

Second, with the three-detector network, we could say for the first time that the polarization structure of the signals was consistent with GR. The fact that the two LIGO detectors are as close to co-aligned as possible (given that they are placed on a sphere and are some 3000 km separated) had meant that no conclusions on the polarization state of previous signals could be made. Virgo has a significant projection on the $45^{\circ}$ $\times$ polarization state, and by considering the quality of fit to the three signals we can say that for GW170814 purely tensor polarization is strongly favored over purely scalar or vector polarizations – consistent with General Relativity. More signals will allow more tight constraints, but the added dimension of Virgo allowed an added dimension of inference from the signal. 

But events of three days later made this first detection even more important. On August 17th, at 12∶41:04 UTC (a very civilized hour for US East Coast scientists for once!), one of the LIGO pipelines -- gstlal -- rang an alert for one of its architects, Chad Hanna, that a very significant single detector trigger from Hanford was seen, with a mass range corresponding to a binary neutron star coalescence. This single-detector trigger would normally be due to a detector noise ‘glitch’ in one instrument and not the other, or a case of only one instrument in observing mode. As we opened up windows to take a closer look, a background task designed to watch for coincidences between gamma ray bursts and GW triggers produced an estimate of a very high significance for this pair of events with a false alarm rate of some $10^{-16}$ Hz.  A handful of colleagues inspected the Livingston trace and discovered the automated software discounted the trigger due to very strong ‘glitch’ in the time series in the middle of the otherwise clear trace. It was quickly determined that a simple algorithm to ‘window out’ the glitch left a signal that could be analyzed with low-latency software, and it was clear that very consistent waveforms had been seen in the two LIGO detectors of very high significance and plausibly associated with the GRB (with some 1.7 seconds interval between the inferred GW arrival time and the GRB arrival time). Attention turned to Virgo: Had they been on the air? Did they also see this signal?

Indeed, Virgo was running at its modest O2 sensitivity; however, the usual data pipeline from the Virgo data to the computers in the US being used for the signal analysis was down, and an alternative was pursued to get the Virgo data together with the LIGO data to systems where the three signals could be combined coherently into a sky map. While this work started, the first machine-readable alert was sent to the electromagnetic and particle groups with which there were agreements for low-level and low-latency data sharing, roughly 27 minutes after the signal traversed the Earth.

The combined LIGO and Virgo data were employed to form a sky map, as had been practiced with GW170814 just three days earlier, and the result quickly checked for plausibility and consistency with the GRB signal seen in both Fermi and INTEGRAL. About 5 hours after the signal was first seen, we sent this skymap out to our partners. The next observatories which had a chance to see the signal were in Chile, where the sun had not yet set, but groups put together their observing plan and almost as soon as the sun set 6 groups, with Swope the first among them, very quickly independently identified a new optical signal near a host Galaxy, within the region supplied in the LIGO-Virgo skymap. A follow-up campaign covering all wavelengths of electromagnetic radiation as well as neutrinos followed, with extremely rich results (see for instance ApJ 848:L12). 

One interesting side note is that in fact, the signal in the Virgo detector from GW170817 is not visible in a spectrogram of the event time. The `antenna pattern' of the laser interferometer gravitational-wave detectors is rather smooth, with best sensitivity on average overhead and underfoot, and a sensitivity of one-half for signals arriving along one arm or the other. But a signal arriving at $45^{\circ}$ to the arms will not excite the differential arm lengths -- and the signal source for GW170817 was very close to that null in response for the Virgo antenna. The sky map effectively uses the detectors as a phased array, and the very small signal in Virgo was just as meaningful as a large signal would have been, and thus Virgo played a central role in this story. We are grateful that Nature provided a ‘test injection’ three days earlier to verify that the Virgo instrument was working and that we had the analysis tools for forming the sky map also tested so recently. 

The O2 Observing run ended just 13 days later, with this `bang'. Since that time, both Virgo and LIGO have been working to improve the sensitivity of the detectors in preparation for O3, planned to start in early 2019. Virgo expects to make a significant step in sensitivity from 26 to 60 Mpc reach for binary neutron star coalescences, and the two LIGO detectors are expected to reach 120 Mpc reach (from 60 and 100 in O2). With this improvement, and remembering that we measure the amplitude of the GW, the rate of detections by the network is expected to increase significantly. The installation work is nearly finished in Virgo and LIGO, and the commissioning starting up, leaving from August 2018 until the start of the run for chasing noise sources and tuning control systems. 

In parallel, the Virgo and LSC teams are preparing for the onslaught of signals expected in O3. Work continues on a generous handful of papers still remaining from the previous runs, in parallel with the development of new systems to provide fully automated low-latency alerts -- every minute counts in alerting the non-GW observing community. A very significant change is that the LIGO and Virgo signals will be sent out as public alerts, consistent with our policy to work toward more engagement of a larger scientific community. For our published papers, we continue to release one hour of data around signals, and are catching up on posting posterior samples along with data behind the figures, again with the hope that there will be a growing group outside of the Collaborations looking at the data. 

O3 is expected to last about one calendar year (and with a chance that the Japanese KAGRA instrument will join near the end), followed by further commissioning to reach the sensitivity foreseen for Advanced LIGO and Advanced Virgo by design. O4 will be performed with that sensitivity, and then in the early 2020’s we expect to make another improvement to the instrument sensitivity – in the case of LIGO, to add frequency-dependent `squeezed light' to reduced quantum noise and to install mirrors with lower thermal noise. This `A+' upgrade should yield a factor of 1.5-2 in sensitivity for observing into the 2020’s. 

Yet further in the future are plans to introduce mild cryogenics in one or more detectors in the existing observatories, along with other techniques that would be used in a new ‘Third Generation’ observatory and instrument. The Gravitational Wave International Committee (\href{https://gwic.ligo.org}{https://gwic.ligo.org}) is helping to coordinate efforts in Europe and in the US to ensure we have a coherent and robust science case, and that we take advantage of the benefits of a global organization for both technology and governance. A success-oriented schedule suggests that we could have instruments a factor of 10 more sensitive (so an event rate a factor of 1000 greater) than Advanced LIGO and Virgo operating by the mid-2030’s.  We hope that will be in conjunction with LISA -- the space-based gravitational-wave antenna -- for both multi-messenger and multi-band science. 

We have lost a number of the early key contributors to the field of interferometric detection of gravitational waves in recent times: Ron Drever, Adalberto Giazotto, Albrecht R\"udiger, and Roland Schilling. They all were able to see the first fruit of their efforts, happily, but we miss them all. With that, and with thanks to them and the thousand plus persons who made this story possible -- 

The future of this new field of gravitational wave observation looks very promising; and, happily, some promises were already fulfilled.

\vfill\eject

\section*{\centerline {Mathematics, Physics, and their Interaction}
\centerline {Conference in Honor of} \centerline {Demetrios Christodoulou's 65th Birthday}}
\addtocontents{toc}{\protect\medskip}
\addtocontents{toc}{\bf Conference Reports:}
\addcontentsline{toc}{subsubsection}{
\it Christodoulou Conference, by Lydia Bieri}
\parskip=3pt
\begin{center}
Lydia Bieri, University of Michigan 
\htmladdnormallink{lbieri-at-umich.edu}
{mailto:lbieri@umich.edu}
\end{center}

This international conference in honor of Demetrios Christodoulou's 65th birthday was held July 10-14, 2017, at ETH Zurich in Switzerland. It brought together mathematicians, physicists and astrophysicists working in general relativity, astrophysics, fluid dynamics and partial differential equations in general. The realm of talks spanned from experiments to mathematical theory. Around 150 participants attended the conference, including experts as well as junior researchers and students. 

Physics and mathematics are interwoven and more than ever interdisciplinary research is required to solve the challenging problems. A culmination of this fact is the theory of general relativity, where not only the Einstein equations govern the laws of the universe but also give them a geometric structure. Demetrios' research reflects this interplay of mathematics and physics in a most beautiful way. 

The talks were spread over 5 days. The first day featured a discussion session after the reception in the evening, which was well-attended and interesting new questions were addressed. There was also a conference dinner, during which several of Demetrios' students and collaborators shared interesting and also amusing stories. 

General relativity (GR) constituted the largest part of the program. Gilbert Weinstein commenced the talks speaking about harmonic maps with prescribed singularities and applications to general relativity. These maps have shown useful in finding lower bounds on the total mass among other things. Gilbert with M. Khuri and S. Yamada use similar ideas to investigate a class of black holes with non-standard topology in 5 dimensions. Mihalis Dafermos reported on the cosmic censorship conjectures. These conjectures, originally put forth by Roger Penrose, remain among the most central unsolved problems in GR. Whereas the strong cosmic censorship conjecture suggests that for generic 
Einstein vacuum initial data, the solution spacetime determined by 
initial data cannot be extended as a suitably regular Lorentzian 
manifold; the weak cosmic censorship conjecture says that for generic asymptotically flat vacuum data, the resulting  vacuum spacetime has a complete 
future null infinity, thus no naked singularities should occur. Obviously, there is room to formulate what ``generic" data should be. Christodoulou's work in the 1980s and 1990s  led him to a more precise formulation of these conjectures. In particular, studying the collapse of a spherically symmetric self-gravitating scalar field for certain classes of initial data, Christodoulou showed that such naked 
singularities may occur, but they are unstable, therefore establishing a proof of a definitive version of cosmic censorship in this case. Dafermos has done work investigating cosmic censorship. In his latest results with J. Luk, they prove that 
one can extend the maximal Cauchy evolution of appropriate Einstein vacuum data across a piece of the Cauchy horizon of a Kerr black hole as a Lorentzian manifold with $C^0$-metric. Their work suggests that a $C^0$-version of strong cosmic censorship is false, but that a more precise version due to Christodoulou may still hold, in the sense that the Cauchy horizon is $C^0$-stable but not more, that is, at the same time it will be singular in the sense suggested by the vacuum weak null singularities. Exciting new insights about cosmic censorship was given by Robert Wald. 
If an extremal or nearly extremal black hole can be made to absorb matter with sufficiently large angular momentum or charge as compared with its energy, one would obtain an apparent contradiction with cosmic censorship. 
Wald presented recent work where no over-charging or over-spinning of a black hole can occur, provided only that the non-electromagnetic matter satisfies the null energy condition, taking into account all second order effects. 

Claudio Bunster explored gravitational domain walls and the dynamics of $G$. A. Shadi Tahvildar-Zadeh presented work with M. Kiessling on how to remedy the problem of infinities inherent in Maxwell-Lorentz electrodynamics of point charges, which were identified by Hermann Weyl as the main obstacle on the path to a possible symbiosis of GR and QM. 
Shing-Tung Yau talked about his crucial work with P. Chen and M.-T. Wang on quasi-local mass. Let us remind ourselves that there are many important physical quantities and questions that 
were understood in Newtonian mechanics, however, their
counterparts in GR are not easy to formulate, let alone to understand. Energy and mass are prominent examples, which are understood for certain classes of spacetimes, but the ultimate insights are still lacking. Yau and collaborators have made important contributions to shed light on this problem. In particular, the Wang-Yau quasi-local mass has good properties and shows fruitful to solve other problems in GR. Recently, Yau and collaborators have extended their work to define quasi-local energy and optimal isometric 
embeddings in reference to the de-Sitter (dS) and the Anti-de-Sitter 
(AdS) spacetimes. Richard Schoen explained recent work with S.-T. Yau on a proof of the positive energy theorem in higher dimensions. Their famous result of 1979 established positivity of total mass for large classes of $4$-dimensional spacetimes. Ed Witten gave a different proof of positivity. Lately, Schoen and Yau considered cases when the dimension is greater than $8$ and the manifold is not spin. The proof is accomplished by extending the minimal hypersurface approach in the presence of singularities and controlling the singular sets which arise. 
Gerhard Huisken reported on various foliations of asymptotically-flat $3$-manifolds 
arising as spacelike slices in Lorentzian spacetimes, modeling isolated gravitating systems. 
These have proven crucial to understand physical concepts such as (quasi-local) mass, and momenta in a geometric way independent of coordinate systems. 

Sergiu Klainerman talked about the monumental work he and Demetrios Christodoulou accomplished in the proof of the nonlinear stability of Minkowski space and the impact it has had on research in GR and in nonlinear wave equations. One main outcome of their work has been a complete understanding of the null asymptotics for physically interesting spacetimes. Based on that knowledge Christodoulou derived the null memory effect of gravitational waves. In a linearized theory, ordinary memory had been known since 1974, when Ya.B. Zel'dovich and A.G. Polnarev computed it. Moreover, Christodoulou's fundamental result of black hole formation of 2008 combined techniques developed in the stability proof with new ones. Mathematically, the methods developed by Christodoulou and Klainerman have proven crucial in many other nonlinear hyperbolic problems. The Christodoulou-Klainerman stability proof was generalized to the Einstein-Maxwell 
case by Nina Zipser in 2000; then in 2007 Lydia Bieri generalized it for the Einstein vacuum equations obtaining borderline decay of the data at infinity, and less regularity. There has been a lot of work in various directions, that was initiated by the Christodoulou-Klainerman result. 
Hans Lindblad reported about a recent result of his and Martin Taylor proving stability for the massive Einstein-Vlasov system in the harmonic gauge. 
Thibault Damour reviewed the theoretical developments on the motion and gravitational radiation of binary black holes that have been crucial in interpreting the LIGO events as being emitted by the coalescence of two black holes. In particular, he presented the effective-one-body (EOB) formalism and how EOB and numerical relativity are put to work to compute templates that have been used to search the first coalescence signals, and to measure the masses and spins of the coalescing black holes. 
Lydia Bieri explained the gravitational wave memory effect and gave new insights. 
GR predicts that gravitational waves change the spacetime permanently, which results in a permanent displacement of test masses, the memory. For a long time, it was believed that there was only one type of memory, and that what Ya. B. Zel'dovich and A. G. Polnarev found in a linearized theory was the ``linear" version of the ``nonlinear" one that D. Christodulou found in 1991 in the fully nonlinear theory. Lydia Bieri and David Garfinkle showed that these are in fact two different memory effects, the former due to fields that do not reach null infinity, the latter due to fields that do reach null infinity. Bieri, Chen and Yau showed that in the Einstein-Maxwell equations a specific component of the electromagnetic field contributes to the null memory, and Bieri and Garfinkle derived the contribution from neutrino radiation to the null memory in GR. Robert Wald and A.Tolish computed interesting examples of ordinary and null memories. It is interesting to see that there is no memory in higher dimensions. See the works by 
Garfinkle, S. Hollands, A. Ishibashi, Tolish and Wald as well as G. Satishchandran and Wald. There has been a wealth of work on gravitational memory and on analogs of memory in other theories. Bieri and Garfinkle for the first time outside of GR derived the two analogs of the memories in electromagnetism for the Maxwell equations. A. Strominger and several collaborators, followed by a large group of researchers, have worked on memory analogs in many other theories. In the cosmological setting, for Einstein-de Sitter the memory is enlarged by a factor involving redshift (Bieri and Garfinkle) and the same holds for Friedmann-Lema\^itre-Robertson-Walker (FLRW) (Tolish and Wald). Bieri, Garfinkle and N. Yunes derived that in a $\Lambda$CDM cosmology with large inhomogeneities the memory in the cosmological zone is multiplied by a factor involving not only redshift but also weak gravitational lensing. It is believed that gravitational wave memory will be detected in the near future. See the result by P.D. Lasky, E. Thrane, Y. Levin, J. Blackman and Y. Chen who suggest to detect memory with LIGO by stacking black hole merger events. 

Barry Barish reported on the major results and perspectives of LIGO. A major breakthrough of GR happened in 2015 with LIGO's first detection of gravitational waves. Since then several events have been recorded by the detectors. This marks the beginning of a new era where we get information directly from the universe itself. The improved LIGO and future detectors will give a much deeper view into the universe.  
Further on cosmology, Ruth Durrer explained that cosmology cannot really be formulated without GR. She gave new insights on the cosmic microwave background (CMB), weak lensing and weak lensing of the CMB. She also explained how planned high precision CMB polarization experiments will be able to measure effects of frame dragging on cosmological scales. David Spergel shared his insights on how to measure the geometry and topology of the universe with CMB observations. In particular, he discussed how precision measurements of the CMB constrain the universe's shape and show that the universe is nearly flat and very large. 
As we know, many interesting mathematical and physical features occur in other theories. 

Wilhelm Schlag reported on longterm dynamics of dispersive evolution equations. In particular, he talked about 
developments dealing with the description of asymptotic states of solutions to semilinear evolution equations, and new results on damped subcritical Klein-Gordon equations. 
Joachim Krieger spoke about results on regularity and singularity formation of critical wave maps and the role that by now classical work by Christodoulou-Shatah -Tahvildar-Zadeh played in their genesis. 
Going further, the half-wave maps equation is a novel geometric evolution equation, which displays many intriguing mathematical properties with links to minimal surfaces, conformal symmetry, and completely integrable systems. 
Enno Lenzmann described how the motivation for this evolution equation comes from exactly solvable quantum systems (Haldane-Shastry models) and completely intergable Calogero-Moser spin systems. He then presented new results of his with P. G\'erard and A. Schikorra about the explicit classification of traveling solitary waves together with their complete spectral analysis. 
Carlos Kenig explained the energy channels method 
and some of its applications to study singularity formation for nonlinear wave equations. 
Sijue Wu talked about water waves modeled by the Euler equations under appropriate assumptions. 
Another field of Christodoulou's research combining mathematics and physics are the Euler equations. 
They form the core part of many fluid models in endless applications. Within mathematics, the Euler equations belong to one of the hot topics and big challenges of modern research. One would like to understand the dynamics, singularities or stability properties of the solutions. These questions are tackled in the framework of the Cauchy problem, where for certain types of initial data and conditions one would like to solve the equations (locally or globally in time) and derive a full description of the solutions. Among the many applications, water waves are described by the Euler equations for an incompressible, inviscid and 
irrotational fluid with air density zero. 
Sijue Wu reported on recent results on 
$2$-dimensional water waves with angled crests. In particular, she explained the 
well-posedness of the Cauchy problem including water waves with non-$C^1$ interfaces. 

Jalal Shatah gave exciting insights on weak turbulence. 
Andrew Majda talked about 
low-dimensional reduced-order models for statistical response and uncertainty quantification in turbulent systems. 
Turbulent dynamical systems characterized by both a high-dimensional phase space and a large number of instabilities are ubiquitous among many complex systems in science and engineering including climate, material, and neural science. 
The nature of the problem is such that there is a rapid growth of small uncertainties from imperfect modeling equations or perturbations in initial values, requiring naturally a probabilistic characterization for the evolution of the turbulent system. 
There are many challenges to overcome. Majda discussed a general mathematical framework to construct statistically accurate reduced-order models that have skill in capturing the statistical variability in the principal directions with largest energy of a general class of damped and forced complex turbulent dynamical systems. 
Zhouping Xin explained recent results on transonic shocks in curved nozzles. In particular, 
he discussed some steady compressible flows in nozzles with variable cross sections. To be precise, he considered 
a nonlinear free boundary value problem with nonlinear boundary conditions for mixed type equations and discussed the existence of single and multiple transonic shocks in terms of the geometry of the nozzle and the given exit pressure. 
Many nonlinear partial differential equations (PDE) arising in mechanics, geometry, and other areas naturally are of mixed elliptic-hyperbolic type. Important examples include shock reflection-diffraction problems in fluid mechanics (the Euler equations) and isometric embedding problems in differential geometry (the Gauss-Codazzi-Ricci equations), among many others. 
Gui-Qiang Chen presented natural connections of nonlinear PDE of mixed elliptic-hyperbolic type with these longstanding problems and discussed some of the most recent developments in the analysis of these nonlinear PDE. He also gave ideas on developing and identifying mathematical approaches, ideas, and techniques for dealing with the mixed-type problems. 
Domenico Giulini reported on aspects of $3$-manifold theory motivated by GR. 

Many speakers shared personal stories on how they have interacted with Demetrios, some of which were most amusing. And Demetrios himself added a few more, among them he explained how working on problems in physics made him become a mathematician. In particular, John Wheeler, his PhD thesis advisor, gave him the following problem in 1968: the formation of black holes in pure GR, by the focusing of incoming gravitational waves. The more Demetrios studied this problem, the more he realized that geometry and PDE play a crucial role, and that the general realm is far more intricate. 
Demetrios then added, smiling, that he solved this problem in full details in 2008; Wheeler had changed the topic of Christodoulou's thesis, so that the latter received his PhD in 1971. Demetrios' 2008 black hole formation result is a masterpiece of physics and mathematics, relying in particular on physical insights combined with new and deep geometric-analytic methods. Someone in the audience pointed out that, being a mathematician and a physicist, Demetrios was indeed quite ``interdisciplinary". 

This conference combined most recent research and deep insights from many highly-active fields, and it sparked new ideas for future research. The participants enjoyed the vibrant discussions within and across the different topics. As an especially nice aspect of this conference emerged the interdisciplinary interactions. 

The conference website can be found at 

\href{https://www.math.ethz.ch/fim/conferences/past-conferences/2017/christodoulou.html}
{https://www.math.ethz.ch/fim/conferences/past-conferences/2017/christodoulou.html}

\vfill\eject
\section*{\centerline
{Stephen Hawking (1942-2018)}}
\addtocontents{toc}{\protect\medskip}
\addtocontents{toc}{\bf Obituaries:}
\addcontentsline{toc}{subsubsection}{
\it  Stephen Hawking (1942-2018), by Robert M. Wald}
\parskip=3pt
\begin{center}
Robert M. Wald, University of Chicago
\htmladdnormallink{rmwa-at-uchicago.edu}
{mailto:rmwa@uchicago.edu}
\end{center}

Stephen Hawking died peacefully in his home in Cambridge, England on March 14, 2018. His life story of making truly groundbreaking discoveries while succumbing to ALS---and his continuing to lead an extremely active and productive life for more than three decades after he completely lost the ability to speak and was nearly totally paralyzed---will be well known to most readers. Further information about Hawking's life is readily available from many sources, including the movie ``The Theory of Everything,'' for which Eddie Redmayne deservedly received the best actor Academy Award for his portrayal of Hawking. I will therefore not attempt to review anything about his life here. I also will not attempt to summarize his truly major contributions to the singularity theorems, the classical theory of black holes, quantum cosmology, and many other areas. However, I would like to make some remarks on his most famous work: the discovery of thermal particle creation by black holes.

I recently carefully reread his paper ``Particle Creation by Black Holes'' (Commun. Math. Phys. {\bf 43}, 199-220 (1975)) in connection with the closing remarks I gave at Hawking's 75th birthday conference in July, 2017. Everyone knows that this is a groundbreaking paper, but all of the reasons why it is such a truly amazing paper may not be as well known. There are many instances of great discoveries in science where the author makes a bold new hypothesis that miraculously turns out to be right. It is all the more remarkable that this is {\em not} such an example. All of Hawking's results in this paper are derived from general relativity and the basic principles of quantum field theory by careful reasoning and calculations, with no room for any additional hypotheses.

A static (``eternal'') Schwarzschild black hole spacetime is relatively simple to consider on account of its time translation symmetry. However, a past event horizon is present in this spacetime, and its presence introduces an ambiguity in the initial conditions for a quantum field. To avoid this ambiguity, Hawking considered instead a spacetime in which gravitational collapse to a Schwarzschild black hole occurs. This removes the ambiguity---and the gravitational collapse spacetime is much more physically relevant to consider in any case---but the lack of time translation symmetry would seem to make intractable the problem of solving for the dynamical evolution of a quantum field. However, Hawking realized that if one considers the {\em backward in time} evolution, a remarkable simplification occurs: A wave that passes very close to the horizon when evolved backward in time gets highly blueshifted, and the geometric optics approximation can be used to evolve it through the dynamical region. 

Particle creation will occur when dynamical evolution results in the mixing of positive and negative frequencies. Thus, the key calculation in the paper consisted of starting with a wave that is purely positive frequency with respect to Killing time in the future and determining its positive and negative frequency parts in the past. In view of the geometric optics behavior, Hawking realized that this amounts to determining the relationship between positive frequency with respect to Killing time and positive frequency with respect to affine time near the horizon of a black hole. Hawking then calculated the positive and negative frequency parts with respect to affine time of a wave with Killing time frequency $\omega$. The form of this expression is not very illuminating except that at the relevant high frequencies the positive and negative frequency parts are very closely related, with the only difference involving a factor containing the logarithm of the affine frequency. Using an analyticity argument, Hawking the showed that the magnitudes of positive an negative parts are simply related by a factor of $\exp(\pi \omega/\kappa)$, where $\kappa$ is the surface gravity of the black hole. This fact is sufficient to calculate the expected number of created particles. However, this number is infinite. Nevertheless, by re-doing the calculations considering wave packets that are nearly of frequency $\omega$ but are localized in time, Hawking showed that this infinity is actually physical---the result of a steady rate of particle creation over all (late) times. When this finite, steady flux of created particles is calculated, the square of the above exponential factor comes in and corresponds precisely to a Boltzmann factor, yielding the astounding result that black holes emit thermal radiation at temperature $T= \kappa/2 \pi$. As Hawking then pointed out, this is just what is needed for the consistency of black hole thermodynamics.

If the paper ended there, it would easily have ranked as one of the most remarkable papers in physics written in the second half of the 20th century. But, in my view, what makes this paper so truly amazing is that Hawking was not content to stop there:

First, Hawking had done the above calculation for a massless scalar field. He then argued that (despite some nontrivial differences) the calculation would continue to hold for electromagnetic and linearized gravitational fields. He then showed that it would also continue to hold for massless Fermi-Dirac fields, but that there would be a sign change because of the nature of the Dirac product versus the Klein-Gordon product, so that one would end up with a Fermi-Dirac distribution rather than a Bose-Einstein distribution. Finally, he argued that the results also would hold for non-zero rest mass. So, one truly gets thermal emission of {\em all} species of particles.

Next, the above calculation had been done for a spherically symmetric spacetime, which is exactly Schwarzschild outside of the collapsing matter. Hawking then argued that, in fact, the late time radiation depends only on the asymptotic final state of the black hole, not the details of the collapse, so his results hold for any collapse that settles down to a Schwarzschild final state. This is a nontrivial argument that occupies about a page of the paper.

Having argued that the late time radiation depends only on the black hole final state, Hawking then derives what the particle creation would be if the final state were a Kerr black hole. This is a very nontrivial generalization on account of superradiance, but Hawking does this in one paragraph. What about charged fields if the final state were a Reissner-Nordstrom black hole? A similar superradiance behavior occurs there, and Hawking shows that this has a similar effect on particle creation.

Hawking then goes on to consider back-reaction of particle creation on the black hole. This was completely new territory. Hawking first argued that an observer freely falling into the black hole should not see any large local effects---thereby making this paper the first anti-firewall paper ever written. He then argued that if there are no large local effects near the horizon, then there must be a flux of negative energy into the black hole corresponding to the positive energy flux at infinity. But this, he argued, would cause the black hole to lose mass and evaporate within a finite time. He then drew a spacetime diagram of an evaporating black hole---a diagram that has been redrawn in literally thousands of papers.

Are there any flaws or errors in the paper? I do not find Hawking's heuristic picture of particle creation involving tunneling to be a good description of the phenomenon, but, as Hawking clearly says, this is just a heuristic picture that is not used to derive any conclusions, so one cannot even call this a flaw. Hawking believed at the time that he wrote that paper that the ambiguities inherent in the definition of ``particles'' near the black hole would give rise to ambiguities in defining the quantum field stress-energy tensor; that the quantum stress energy tensor would therefore have a status similar to that of a gravitational stress-energy pseudotensor; and that therefore one could talk meaningfully only about its averaged/global effects. This turned out not to be the case---the quantum stress-energy tensor is a perfectly good quantum field observable. However, the only effect of Hawking's cautious treatment of the quantum stress-energy tensor was to make his arguments on back-reaction more awkward and difficult to make; all of the arguments are correct. Thus, after careful scrutiny of the paper 43 years after it was written, the only genuine errors I have been able to find are that the word ``gauge'' is misspelled in 4 places, and the name ``Bekenstein'' also is misspelled.

It is a truly amazing paper, written by an even more amazing person!

\vfill\eject
\section*{\centerline
{Joseph Polchinski (1954-2018)}}
\addtocontents{toc}{\protect\medskip}
\addcontentsline{toc}{subsubsection}{
\it  Joseph Polchinski (1954-2018), by Gary Horowitz}
\parskip=3pt
\begin{center}
Gary Horowitz, University of California, Santa Barbara
\htmladdnormallink{gary-at-physics.ucsb.edu}
{mailto:gary@physics.ucsb.edu}
\end{center}

Theoretical physicists lost one of their most brilliant colleagues  when Joe Polchinski  passed away on February  2, 2018, after a two year battle with brain cancer.   Joe was a high energy theorist who made fundamental contributions to quantum field theory and string theory. He literally ``wrote the book" on string theory, publishing a widely used two volume set on the subject in 1998. But for gravitational physicists, he is best known for two contributions that directly apply to black holes.

The first contribution started with a discovery that Joe made (with two graduate students)  that, contrary to widespread belief, string theory is not just a theory of strings \cite{Dai:1989ua}. There are also higher dimensional extended objects he called D-branes. (The D stands for Dirichlet, and refers to the boundary conditions at the ends of the string.) These objects had been missed earlier because they are nonperturbative, having a tension that is proportional to $1/g$ where $g$ is the string coupling constant. Joe later showed that these objects carry charges \cite{Polchinski:1995mt} and are closely related to  the extended charged black holes (or ``black branes") that Andy Strominger and I had found a few years earlier \cite{Horowitz:1991cd}. More precisely, D-branes can be viewed as the source of the extremal limit of these objects. By combining these D-branes  in the right way, Andy and Cumrun Vafa were able to count the microstates of an extremal black hole for the first time, and reproduce the Hawking-Bekenstein entropy \cite{Strominger:1996sh}. 
Joe's D-branes turned out to have many other applications as well, and led directly to the discovery of gauge/gravity duality which has dominated work in string theory for the past two decades.

In 2012, Joe touched off an explosion of interest in the black hole information puzzle.  Working with  Don Marolf and two graduate students, he extended earlier arguments by Mathur \cite{Mathur:2009hf}  and showed that three widespread beliefs about black holes were inconsistent \cite{Almheiri:2012rt}. One cannot maintain that nothing happens to an observer that falls into a large black hole, that black hole evaporation is a unitary process, and that ordinary local quantum field theory holds outside the horizon of a macroscopic black hole. They raised the possibility that one might have to give up the first belief and suggested there might be a ``firewall" near the horizon of a black hole that has evaporated down to half its original size. Although this suggestion remains  controversial, it has stimulated many new ideas which are being pursued today. 

Joe was an avid bicyclist and would often lead visitors to the KITP on strenuous rides up the mountains behind Santa Barbara. As our chancellor said at the time of his death: ``Joe was known for climbing mountains, both intellectual and literal. Many of our colleagues fondly remember epic bike rides with him to the top of Gibraltar and Old San Marcos Pass, discussing life and physics all the way. He set his sights high and navigated fearlessly over all obstacles in order to achieve the extraordinary, encouraging and inspiring others along the way."


Joe was a wonderful colleague and friend. He  gave generously of his time to anyone who asked: students, postdocs, and colleagues.  He will be deeply missed.

\vskip 1cm

\bibliographystyle{JHEP}
\bibliography{all}
  
\end{document}